# Spin-Orbit Torque Driven Chiral Domain Wall Motion in Mn$_3$Sn

Zhengde Xu[1,2,3], Yue Zhou[3,4], Xue Zhang[1,2,3], Yixiao Qiao[1], Zhuo Xu[1] and Dingfu Shao[5†], Zhifeng Zhu[1,6†]

[1]School of Information Science and Technology, ShanghaiTech University, Shanghai, China 201210

[2]Shanghai Institute of Microsystem and Information Technology, Chinese Academy of Sciences, Shanghai, China 200050

[3]University of Chinese Academy of Sciences, Beijing, China 100049

[4]Shanghai Institute of Applied Physics, Chinese Academy of Sciences, Shanghai, China 201800

[5]Key Laboratory of Materials Physics, Institute of Solid State Physics, HFIPS, Chinese Academy of Sciences, Hefei, China 230031

[6]Shanghai Engineering Research Center of Energy Efficient and Custom AI IC, Shanghai, China 201210

**Abstract**

Noncollinear chiral antiferromagnets, such as Mn$_3$X (X = Sn, Ge), have garnered significant interest in spintronics due to their topologically protected Weyl nodes and large momentum-space Berry curvatures. In this study, we report rapid chirality domain-wall (CDW) motion in Mn$_3$Sn, driven by spin-orbit torque at over 545.3 m·s$^{-1}$ under a remarkably low current density of 9×10$^{10}$ A·m$^{-2}$. The results demonstrate that the chirality of the domain wall and the direction of the current collectively determine the displacement direction of the CDW. Theoretically, we provide an analysis of the effective field experienced by the octupole moment, uncovering the underlying motion mechanism based on the unique profile of the chiral spin structure. Notably, CDWs with opposite chirality can form within the same Dzyaloshinskii–Moriya interaction sample, and the Néel-like CDW type is dictated by the orientation of the kagome plane rather than the negligible magnetostatic energy associated with the small magnetization (approximately $3.975 \times 10^{-3}$). Additionally, the CDW, with a considerable width of 770 nm, is segmented into three 60° portions due to the six-fold anisotropy in $Mn_3Sn$. These emphasize that CDW motion in $Mn_3Sn$ cannot be quantitatively studied using ferromagnetic frameworks. We also demonstrate that a small external field can effectively regulate CDW velocity. Our comprehensive results and theoretical analysis provide crucial guidelines for integrating antiferromagnet CDWs into functional spintronic devices.

**Introduction**

The concept of "race-track storage", proposed over the past decade, has sparked significant interest in the electrical manipulation of magnetic domain wall (DW) motion [1-5]. DW exhibits remarkable robustness and potential for next-generation memory devices. Utilizing spin-transfer torque (STT) or spin-orbit torque (SOT), ferromagnetic (FM) DW motion can be electrically manipulated at speeds of hundreds of meters per second, although this requires high current densities of up to $10^{12}$ A·m$^{-2}$ [6,7]. In contrast, the velocity of ferrimagnetic DWs can be significantly improved to 5.7 km·s$^{-1}$ due to antiferromagnetic exchange coupling, although it still necessitates large current densities [8-11].

To achieve high-capacity racetrack memory, closely packed DWs are necessary. However, memory density is often limited by the magnetostatic interaction caused by the stray field of neighboring domains. Unlike FMs, antiferromagnets (AFMs) have magnetic moments that are compensated on an atomic scale, resulting in zero stray fields [12-16]. This property makes AFMs favorable for miniaturization and high-density integration, though it also presents challenges for the efficient manipulation and detection of AFM DWs [17,18]. Recent research has seen substantial advancements in the area of non-collinear chiral AFMs, particularly in Mn$_3$X (X = Sn, Ge) systems [19-26]. These materials have demonstrated large distinctive responses, including the anomalous Hall effect [15,27], anomalous Nernst effect [28], and asymmetric magnetoresistance [29], allowing efficient observation of domain structure in nanostructures. This progress opens new avenues for using AFMs as active elements in spintronic devices. Previous research has demonstrated field-assisted [29,30] and STT-driven DW motion [31]. To further improve energy efficiency and advance next-generation magnetic solid-state memory devices, SOT has gained significant attention [25,32-36]. Despite these advancements, SOT-driven chiral domain wall (CDW) motion in non-collinear AFM systems has not yet been reported.

In this work, we provide an in-depth understanding of the profile of CDWs with different chirality. Based on the rotation direction of the cluster magnetic octupole moment, the chirality is categorized into anti-clockwise and clockwise, both of which can be formed in the same sample. This behavior contrasts with FMs, where chirality is determined by the sign of the Dzyaloshinskii–

Moriya interaction (DMI). This unique characteristic offers a novel approach to advancing CDW studies [37]. The scale and structure of these CDWs are consistent with previous experimental observations and theoretical predictions [38-40]. Importantly, we achieve SOT-driven CDWs motion in both chirality systems with a small current density ($9\times10^{10}$ A·m$^{-2}$), accelerating the speed up to 545.3 m·s$^{-1}$ in Mn$_3$Sn/heavy-metal (HM) bilayers. This velocity surpasses typical FM systems, even at lower current densities [41,42]. The direction of motion is determined by the interplay between current and CDW chirality, consistent with our analysis that the motion is cooperatively facilitated by the damping-like SOT effective fields ($\mathbf{H}_{DL}$) and the vertical component of the exchange field ($\mathbf{H}_\perp$). Based on our results, it is worth noting that conventional FM frameworks are insufficient for quantitatively describing CDW motion in Mn$_3$Sn. We analyze this discrepancy from several perspectives. In Mn$_3$Sn, CDW type is dictated by the orientation of the kagome plane, not the competition between magnetostatic energy ($E_{mag}$) and other energy components as in FMs. Additionally, chirality is governed by initial formation conditions, allowing both clockwise and anti-clockwise CDWs to form under the same DMI constant. The unique triangular spin structure of Mn$_3$Sn, with six stable states, divides the CDW profile into three segments, leading to a larger CDW width than FM. These differences highlight the key distinctions arising from the intrinsic properties of Mn$_3$Sn. We also reveal an efficient control of CDW velocity by combining a small external field ($\mathbf{H}_{ext}$) in the kagome plane. These findings offer new insights into controlling CDW dynamics and advancing spintronics using AFMs.

**Methodology**

To study the CDWs, we utilized the atomistic model to construct a nanoribbon Mn$_3$Sn sample [43-45], with a length of 5.7 μm. The energy of this system is determined by the following Hamiltonian,

$$E = A\sum_{i,j} \mathbf{m}_i \cdot \mathbf{m}_j + \sum_{i,j} \mathbf{D}_{ij} \cdot (\mathbf{m}_i \times \mathbf{m}_j) - \sum_i (\mathbf{K}_i \cdot \mathbf{m}_i)^2 - \mu_0 M_s \sum_i (\mathbf{m}_i \cdot \mathbf{H}_{ext}),$$

where the exchange interaction constant $J$ = 9 meV [46,47], the DMI constant $D$ = 0.833 meV [12,20], and the magnetic anisotropy constant $K$ = 0.196 meV [12,20]. The magnetic moment $m_s$ = 3$\mu_B$ [12,20]. The easy axis of crystalline anisotropy points to the nearest Sn atoms (the silver ball shown in Fig. 1(a)). The last term is the Zeeman energy due to the external field $\mathbf{H}_{ext}$.

The dynamics of magnetic moments are described by the coupled Landau–Lifshitz–Gilbert–Slonczewski (LLGS) equations [48-50],

$$\frac{d\mathbf{m}_i}{dt} = -\gamma \mathbf{m}_i \times \mathbf{H}_{eff,i} + \alpha \mathbf{m}_i \times \frac{d\mathbf{m}_i}{dt} - \gamma \frac{\hbar \theta_{SH} J_c}{2eM_s t} \mathbf{m}_i \times (\mathbf{m}_i \times \boldsymbol{\sigma}_i).$$

The first term on the right-hand side represents the precession of the magnetic moment around the effective magnetic field $\mathbf{H}_{eff,i} = -\frac{1}{m_s}\frac{\partial E}{\partial \mathbf{m}}$. The second term describes the Gilbert damping, and the last term is the SOT. $\gamma$ is the gyromagnetic ratio, and $\alpha$=0.003 is the damping constant. The spin-Hall angle $\theta_{SH}$ = 0.06 is used in the simulation [12,20].

**Results and Discussion**

In equilibrium, Mn$_3$Sn has six stable states, each differing by 60° within the kagome planes. As shown in Fig. 1(b), we define them as states a, b, c, d, e, and f, respectively. Each set of three spin sublattices, which takes ferroic order on the Kagome lattice, forms a cluster magnetic octupole.

Importantly, there is a very small non-zero net magnetization ($\mathbf{m} = \frac{\mathbf{m}_1 + \mathbf{m}_2 + \mathbf{m}_3}{3}$), depicted as brown arrows, which is largely parallel to the cluster magnetic octupole moment [39,51]. To study the SOT-driven CDW motion, a Mn$_3$Sn/HM bilayer is constructed with two opposite states positioned at the ends of the sample along the longitudinal direction. As shown in Fig. 1(c), a 180° CDW forms at the interface between two domains, where the magnetic moments transition from the b state to the e state. A current flowing through the bottom HM layer generates a pure spin current by the spin-Hall effect (SHE) and the resulting spin polarization $\boldsymbol{\sigma}$ is perpendicular to the kagome plane. The CDW is formed by the gradual rotation of each triangular spin unit, manifesting as a twisted $\mathbf{m}$ in the kagome plane in a Néel-wall configuration. In conventional perpendicular magnetized FM films, Bloch walls typically form due to balanced $E_{\text{mag}}$, while large DMI tends to convert Bloch walls into Néel walls. For in-plane FM systems, as the sample thickness $t$ decreases, the $E_{\text{mag}}$ becomes more dominant. When $t$ approaches the DW width, the Bloch wall, with its out-of-plane magnetization, generates high $E_{\text{mag}}$ and becomes unstable. To maintain a lower energy state, the DW transitions to a Néel wall, where the magnetization rotates parallel to the film surface, effectively reducing the influence of the $E_{\text{mag}}$ [52,53]. However, in the case of Mn$_3$Sn, the finite anisotropy within the kagome plane constrains the rotation of $\mathbf{m}$ in this plane. Additionally, the small magnitude of $\mathbf{m}$ in Mn$_3$Sn, approximately $\|\mathbf{m}\| \approx 3.975 \times 10^{-3}$, leads to negligible $E_{\text{mag}}$, meaning that energy is not the primary factor governing CDW type. Consequently, the orientation of the kagome plane distinguishes between Bloch-like and Néel-like CDWs, rather than energy considerations. Interestingly, as illustrated in Fig. 1(d) and Fig. 1(e), there are two possible configurations distinguished by the profile of the $\mathbf{m}$ within the CDW: anti-clockwise and clockwise.

The chirality of the CDW corresponds to a clockwise or anticlockwise rotation of **m** when viewed from left to right. For instance, if it consists of states b, a, f, and e (Fig. 1(d)), the **m** rotates anti-clockwise. Conversely, if the CDW structure includes states b, c, d, and e (Fig. 1(e)), the **m** rotates clockwise, defining it as a clockwise CDW. The origin of chirality in Mn$_3$Sn is dictated by initial formation conditions, while in FM, it is determined by the sign of the DMI constant to minimize the DMI energy $\mathbf{H}_{\mathrm{DMI}} = \mathbf{D}_{12} \cdot (\mathbf{m}_1 \times \mathbf{m}_2)$. In addition, from the CDW outlined by the blue and red gradient band in the figure, it can be estimated that the CDW width is 770 nm in the micrometer scale, which is in good agreement with previous experimental and simulation estimations [29,54]. However, estimating the CDW width using the formula $W = \pi\sqrt{A/K}$ for FM yields a width of only 12.05 nm [41,53], which is significantly smaller than the 770 nm observed in our results. This suggests that the triangular spin structure of Mn$_3$Sn requires a longer spatial extent for smooth rotation within the CDW. In terms of the internal profile of the CDWs, unlike conventional FM systems with a single easy axis, either perpendicular or in-plane, **m** exhibits sixfold degeneracy. This causes the CDW to divide into three 60° segments, separated by stable states [29]. As a result, two small plateaus formed, as indicated by the red points in Figs. 1(d) and 1(e).

In thin FM materials, DWs generally tend to form into Néel walls, which can be driven by the SHE. Similarly, we achieved CDW motion driven by the SHE originating from the underlying HM layer. A current density of $J_c = 5 \times 10^{10}$ A·m$^{-2}$ was applied along the +**y** direction in the sample with an anti-clockwise CDW at its center. As shown in the snapshot of Fig. 2(a), the white band representing the CDW moves in the direction opposite to the applied current $\mathbf{J}_c$. When the current direction is reversed, as shown in Fig. 2(b), the CDW motion direction also reverses. This indicates that $J_c$ directly determines the direction of CDW motion. Furthermore, we observe that the initial chirality of the CDW also plays a decisive role. As shown in Fig. 2(c) and Fig. 2(d), the CDW with clockwise chirality, illustrated by the blue and red bands, moves in the same direction as $J_c$, regardless of whether $J_c$ is along the +**y** or −**y**. This indicates that the current direction and the initial chirality of the CDW jointly affect the direction of its displacement. Furthermore, we can determine the profile and chirality of the CDW based on the direction of the current and its motion. Corresponding to the snapshots above, we show the $\varphi$ at each position of the sample at different times in Figs. 2(e)-(h). The center of the sample is defined as 0 nm. The equal spacing between each line indicates uniform motion, and the shape of the CDW remains consistent during its motion, with no noticeable shrinkage or expansion. In addition, the velocity of CDW motion in $Mn_3Sn$ easily surpasses that of typical FM systems, where speeds are generally around 100 m·s$^{-1}$ [42,55]. Recent studies have shown that STT-driven CDW motion can also reach several hundred meters per second and have attributed this to the large nonadiabatic torque [31]. The stable characteristics and effective SOT manipulation of the CDW in non-collinear AFM are advantageous for potential device-level applications.

Fig. 3(a) conceptually illustrates the mechanism of CDW motion. Although the neighboring atoms are antiferromagnetically coupled, **m** is often treated as a collective representation of three triangular spin units. Therefore, by orthogonally decomposing the effective FM coupling exchange field ($H_{ex}$, green arrows) between neighboring **m**, the vertical component ($H_\perp$) on the left side of CDW assists the $H_{DL}$ in reorientation **m**. Conversely, the $H_\perp$ on the right side compensates for $H_{DL}$, limiting the expansion of the CDW on that side. Additionally, because of the strong in-plane magnetic anisotropy that keeps spins restricted in the Kagome plane, the propagation of the CDW is consistently restricted in the **y-z** plane, behaving as a Néel-wall type and moving along the wire from right to left. Consequently, the motion direction depends on the combination of $\mathbf{H}_{DL}$ and $\mathbf{H}_{ex}$. When the chirality of the CDW is reversed to a clockwise fashion, the beneficial $\mathbf{H}_\perp$ shifts to the right side, aligning with the direction of $\mathbf{H}_{DL}$. The cooperation of these two torques results in the CDW moving in the +**y** direction, i.e., the direction of $\mathbf{J}_c$. Our analysis explains the simulation results and guides the direction of CDW motion. It is essential to emphasize that **m** should only serve as a qualitative descriptor of the direction of CDW motion, bearing in mind that the DW type, chirality, width, internal profile, and motion speed of $Mn_3Sn$ differ significantly from those in FM systems. These distinctive characteristics highlight why quantitatively analyzing CDW motion by directly analogizing it to FM systems is inaccurate. The distinct behaviors observed in our results reflect the exceptional physical properties of $Mn_3Sn$.

To further investigate the CDW motion, we present the CDW velocity ($\mathbf{v}_{CDW}$) as a function of $\mathbf{J}_c$, where $\mathbf{v}_{CDW}$ is defined as the displacement of the CDW per unit time. Fig. 4(a) shows the $\mathbf{v}_{CDW}$ for right-to-left propagation under positive current and anti-clockwise chirality. We observe that the current-induced SOT can accelerate the CDW up to 545.3 m·s$^{-1}$ with a current density of only $9\times10^{10}$ A·m$^{-2}$. In addition, $\mathbf{v}_{CDW}$ increases nearly linearly with current density. As current density continues to increase, the CDW motion abruptly disappears and is replaced by the oscillations throughout the sample. For instance, as shown in Fig. 4(b), the time evolution of a **m** located 2000 atoms away from the initial CDW center is depicted under $J_c = 1.3\times10^{11}$ A·m$^{-2}$. This behavior is in the same manner as previous research on the $Mn_3Sn$ switching, which reported oscillation regions under large currents [19,56].

In addition to SOT, applying a small $H_{ext}$ can also regulate the velocity of CDW motion. Based on the analysis above, the $H_{DL}$ generated by $J_c$ along +y direction tends to rotate **m** anti-clockwisely, regardless of the CDW chirality. However, the trajectory varies with the chirality of the CDW. Fig. 4(c) shows the trajectory of the **m** positioned along the path of the CDW motion when $J_c$ is $+5\times10^{10}$ A·m$^{-2}$. For the anti-clockwise CDW (red line), the initial position is marked by a star at −425 nm, and for the clockwise CDW (blue line), it is marked by a triangle at +425 nm. Although both trajectories rotate anticlockwise, their paths along the **y** direction are opposite. These rotation directions align with the analyzed $H_{DL}$, offering valuable insights for regulating CDW displacement. As seen in Fig. 4(d), when a small $H_{ext}$ is applied along the +**y** direction, the $v_{CDW}$ changes significantly. The increase or decrease in velocity depends on whether $H_{ext}$ assists or hinders the rotation of the **m**. For an anti-clockwise CDW (red lines), the **m** rotates from +**y** to −**y**. An $H_{ext}$ pointing to −**y** can therefore accelerate this rotation, thereby increasing the $v_{CDW}$. Conversely, for a clockwise CDW (blue lines), the situation is reversed. The $v_{CDW}$ can be nearly doubled with the assistance of an $H_{ext}$ as small as 100 Oe. This dual regulation by current and $H_{ext}$ opens new possibilities for controlling CDW dynamics. Due to its excellent plasticity, Mn$_3$Sn is anticipated to be a superior material platform for mimicking artificial synapses [57] and developing various future spintronic devices [58-60].

**Conclusion**

In conclusion, we have demonstrated fast SOT-driven CDW in Mn$_3$Sn. Our comprehensive atomic-scale study reveals that CDW displacement is influenced by both CDW chirality and current direction. Theoretically, we analyzed how the $\mathbf{H}_{ex}$ and $\mathbf{H}_{DL}$ guide CDW motion. We argue that analogies between CDW motion in Mn$_3$Sn and FM systems are quantitatively inaccurate due to differences in Néel wall formation, chirality, CDW width internal profile, and motion speed. Notably, the $\mathbf{v}_{CDW}$ easily exceeds 545.3 m·s$^{-1}$, surpassing that of FM DWs, and increases linearly with current density. However, when $J_c$ increases significantly and reaches the oscillation region, CDW motion ceases and is replaced by coherent oscillation. Additionally, the introduction of a small $\mathbf{H}_{ext}$ allows for precise modulation of $\mathbf{v}_{CDW}$. Based on the effective and flexible manipulation method proposed in our study, the merits of having no stray magnetic fields and high CDW velocity make this approach very promising for the next generation of DW-based spintronic devices. The unique characteristics of Mn$_3$Sn CDWs and their effective manipulation highlight the potential of this approach for next-generation domain-wall-based spintronic devices.

†Corresponding Authors: zhuzhf@shanghaitech.edu.cn, dfshao@issp.ac.cn

The data that support the findings of this study are available from the corresponding author upon reasonable request.

**Acknowledgments**: We acknowledge the support from the National Key R&D Program of China (Grant No. 2022YFB4401700), National Natural Science Foundation of China (Grants Nos. 12104301), and Program of China Scholarship Council (Grants No. 202408310265). The simulation conducted in this work is supported by SIST Computing Platform at ShanghaiTech

**Figures**

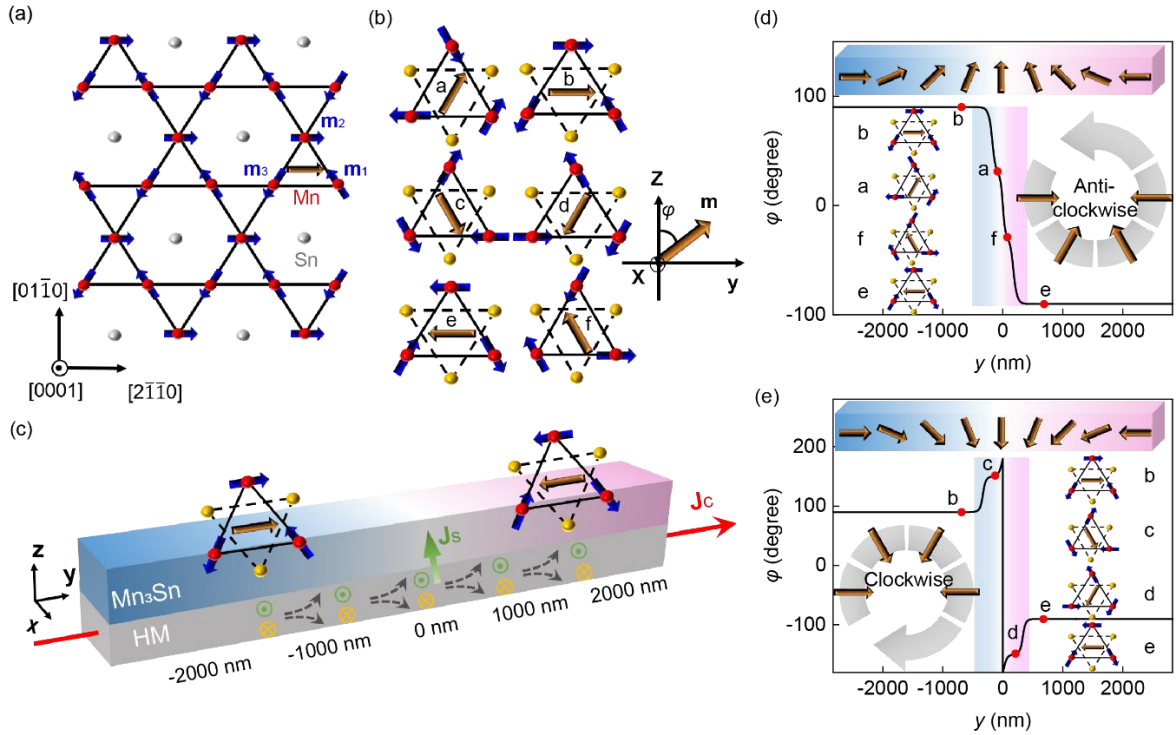

Fig. 1. (a) Atomic structure of Mn$_3$Sn. The Sn (silver circles) and Mn (red circles) atoms lie in the (0001) plane. (b) Six types of cluster **m** in Mn$_3$Sn, which are defined as state a, b, c, d, e, and f, respectively. The corresponding angles are $\varphi$=30°, 90°, 150°, 210°, 270°, 330°. **m** is illustrated by brown arrows. (c) The structure of SOT-driven CDW motion in the Mn$_3$Sn/HM bilayer. (d) Illustration of CDW with (d) anti-clockwise chirality and (e) Clockwise chirality. The insets in (d) and (e) show the schematic diagram of the **m** state in the CDW.

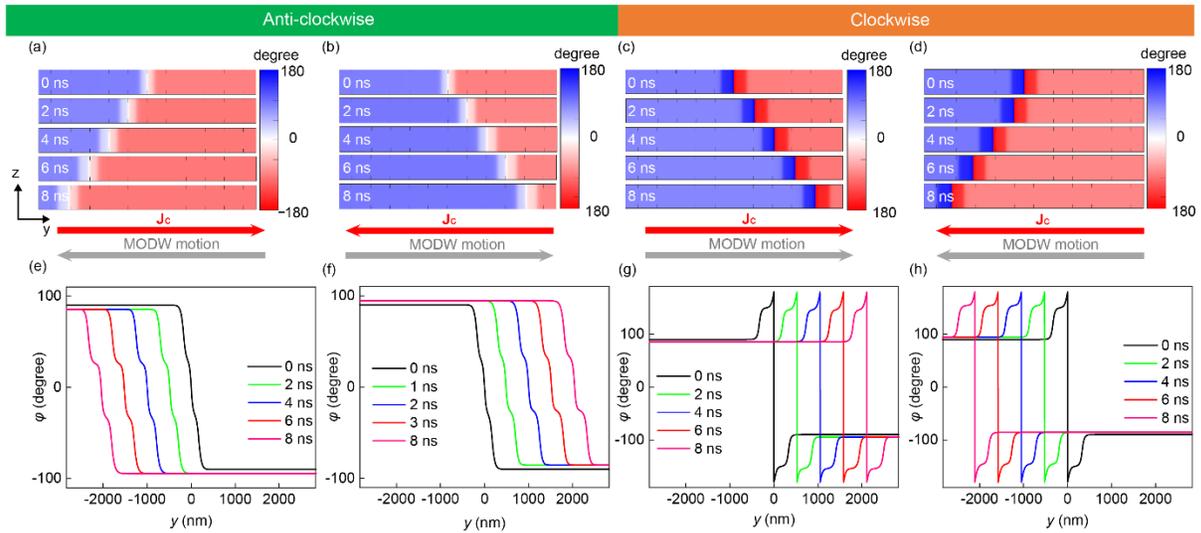

Fig. 2. (a-d) Time evolution of the CDW motion in AFM materials, showing both anti-clockwise (a and b) and clockwise (c and d) chirality. The horizontal axis represents the position along the nanotrack. The red arrows on the bottom indicate the current direction with $J_c=5\times10^{10}$ A·m$^{-2}$ and the gray arrows indicate the direction of the CDW motion. (e-h) Displacement of CDW at different times. The figures are corresponding to (a-d).

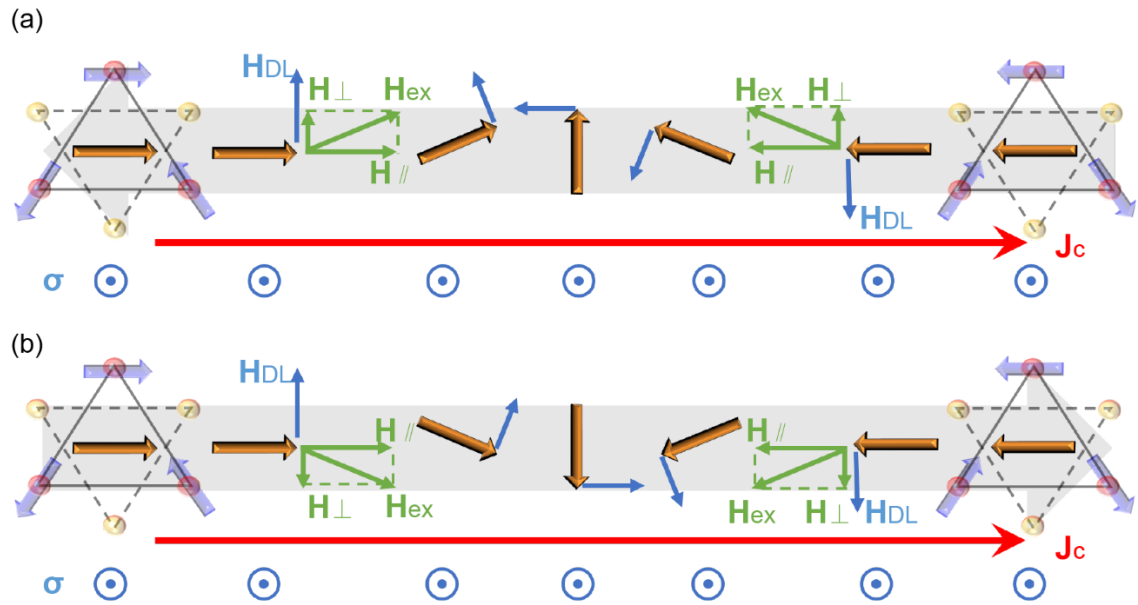

Fig. 3. The mechanism of CDW motion in the system with (a) anti-clockwise chirality and (b) clockwise chirality. Blue arrows represent the effective fields from the SOT, and green arrows represent the $\mathbf{H}_{ex}$.

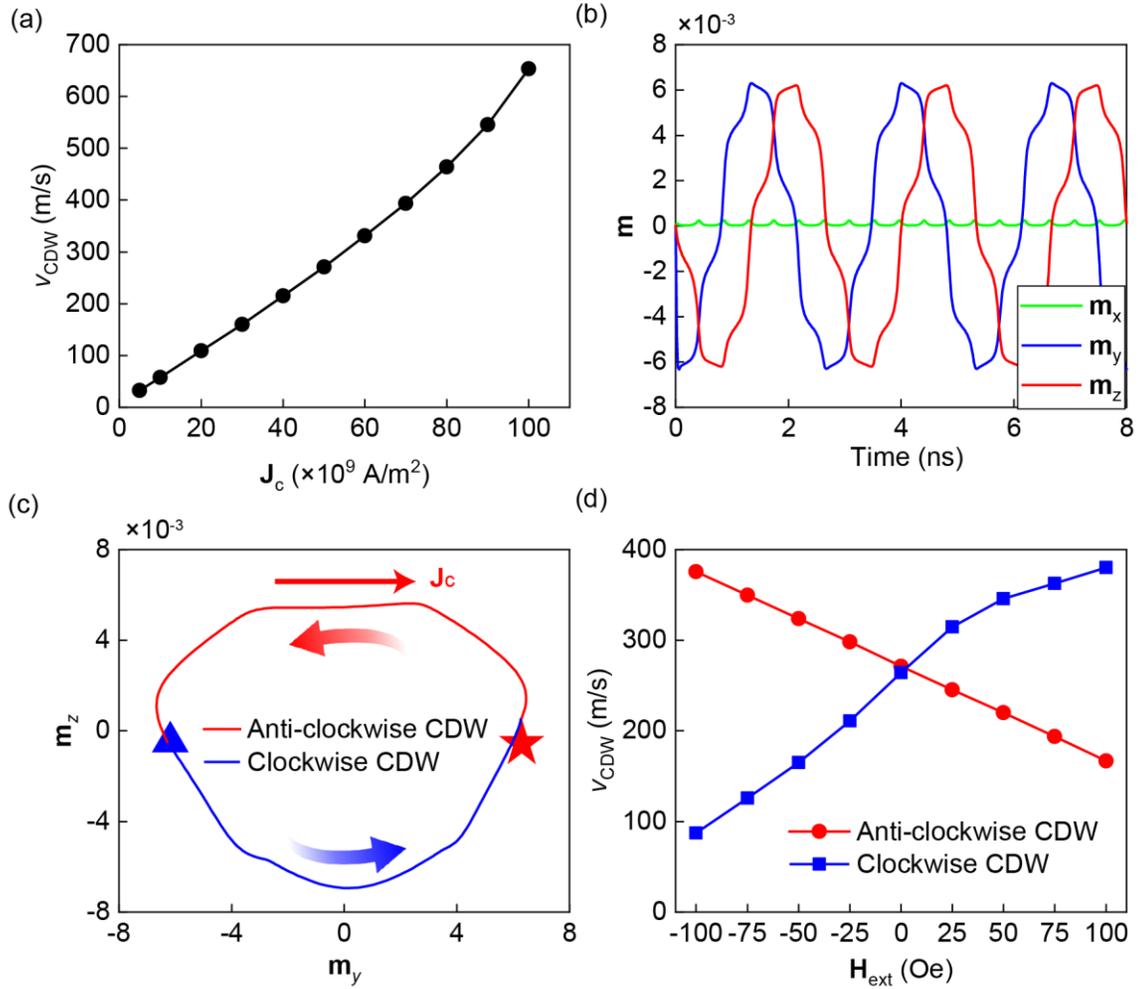

Fig. 4. (a) The velocity of steady CDW motion as a function of $J_c$. (b) Oscillation dynamics of the **m** under large $J_c = 1.3 \times 10^{11}$ A·m$^{-2}$. (c) Cross-sectional view of the **m** trajectory at positions –425 nm (initial point shown as a star) and +425 nm (initial point shown as a triangle) under +$J_c$. (d) $v_{CDW}$ as a function of $H_{ext}$.